\documentclass[twocolumn,showpacs,preprintnumbers,amsmath,amssymb,pra]{revtex4}

\usepackage{epsfig}
\usepackage{graphics}
\usepackage{graphicx}

\usepackage{bm}

\newcommand{\be}{\begin{equation}}
\newcommand{\ee}{\end{equation}}
\newcommand{\bea}{\begin{eqnarray}}
\newcommand{\eea}{\end{eqnarray}}
\newcommand{\ket}[1]{| #1 \rangle}

\begin{document}

\title{Freezing a Coherent Field Growth in a Cavity by Quantum Zeno Effect}
\author{J. Bernu$^1$}
\author{S. Del\'eglise$^1$}
\author{C. Sayrin$^1$}
\author{S. Kuhr$^1$}
\altaffiliation{Present address: Johannes Gutenberg Universit\"{a}t,
Institut f\"{u}r Physik, Staudingerweg 7, D-55128 Mainz, Germany}
\author{I. Dotsenko$^{1,2}$}
\author{M. Brune$^1$}
\email{brune@lkb.ens.fr}
\author{J. M.~Raimond$^1$}
\author{S. Haroche$^{1,2}$}
\affiliation{$^1$Laboratoire Kastler Brossel, Ecole Normale Sup\'erieure, CNRS, Universit\'e P. et M. Curie,
24 rue Lhomond, F-75231 Paris Cedex 05, France\\
$^2$Coll\`ege de France, 11 Place Marcelin Berthelot, F-75231 Paris
Cedex 05, France}

\date{\today}

\begin{abstract}
We have frozen the coherent evolution of a field in a cavity by repeated
measurements of its photon number. We use circular Rydberg atoms dispersively
coupled to the cavity mode for an absorption-free photon counting. These
measurements inhibit the growth of a field injected in the cavity by a
classical source. This manifestation of the Quantum Zeno effect illustrates the
back action of the photon number determination onto the field phase. The
residual growth of the field can be seen as a random walk of its amplitude in
the two-dimensional phase space. This experiment sheds light onto the
measurement process and opens perspectives for active quantum feedback.
\end{abstract}

\pacs{
 03.65.Ta,  
 03.65.Xp,   
 42.50.Pq 
 }

\maketitle

The `Quantum Zeno' effect is a spectacular manifestation of the measurement
back action. The coherent evolution of a quantum system is inhibited by
repeated projective measurements performed at short time intervals. This
effect, theoretically discussed in \cite{zenotheo,Home,Rauch}, has been
observed on the evolution of two-level systems such as molecules
\cite{zenomolec}, trapped ions \cite{zenoion}, cold atoms \cite{zenobec} as
well as on polarization of light \cite{zenopolar}. We describe here the
observation of the Zeno effect on a harmonic oscillator. The build-up of a
coherent field in a high-$Q$ superconducting cavity coupled to a classical
source is inhibited by watching repeatedly the photon number. The measurements
use circular Rydberg atoms as non-destructive probes \cite{lifedeath,
collapse}. The system free evolution is here a run-away coherent classical
process, instead of the two-level quantum dynamics considered in
\cite{zenomolec,zenoion,zenobec,zenopolar}. When watching the system, we record
single trajectories and the field state evolution is obtained by a statistical
analyzis of these data. This study sheds light on the relationship between the
Zeno effect and the measurement back action.

We first revisit the basics of the Zeno effect in the context of our
experiment. A classical source is resonantly coupled to the cavity. The
evolution due to the source--cavity coupling during a time $T$ is described by
the displacement operator $D(\alpha)=\exp(\alpha a^\dagger-\alpha^* a)$
\cite{exploring}, where $\alpha=\lambda T$, $a$ ($a^\dagger$) are the photon
annihilation (creation) operators and $\lambda$ is the complex amplitude of the
source. At this stage, we neglect relaxation, assuming  $T\ll T_c$, where $T_c$
is the cavity damping time. The initially empty cavity contains at time $T$ the
coherent state $\ket{\alpha}=D(\alpha)\ket{0}$ ($\ket{0}$ is the vacuum). The
average photon number, $\overline{n}=|\alpha|^2=|\lambda|^2 T^2$, runs away
quadratically with $T$. This evolution can be split into small steps of
duration $\Delta t\ll T$. The final displacement results from the coherent
accumulation of successive injection pulses, each being described by the
translation operator $D(\lambda \Delta t)$.

We now watch the field evolution. In order to separate the field probing from
its coupling with the source, we alternate measurements with injection pulses
of duration $\Delta t$, such that $|\lambda| \Delta t\ll 1$. After the first
pulse, the cavity contains a coherent state $\ket{\lambda \Delta t}\approx
\ket{0}+\lambda\Delta t\ket{1}$ ($\ket{1}$ is the one-photon state). With a
probability $p_0\approx1-(|\lambda|\Delta t)^2$, the first measurement projects
this state back onto $\ket{0}$. After $N$ iterations of the
injection/measurement sequence, the field is left in $\ket{0}$ with a
probability $p_0^N\approx[1-(|\lambda|\Delta t)^2]^N\approx 1-|\lambda|^2 T
\Delta t$, where $T=N\Delta t$ is the total injection time (we assume that
$\Delta t$ is chosen small enough so that $|\lambda|^2 T \Delta t\ll 1$). The
average photon number, $\overline{n}=|\lambda|^2 T \Delta t$, grows linearly
with $T$. This is strikingly different from the quadratic growth obtained
without measurements. The final mean photon number is smaller than $|\lambda|^2
T^2$, the value reached without measurements. The corresponding reduction
factor, $T/\Delta t$, can be made arbitrarily large. At the limit of infinitely
many measurements ($\Delta t\rightarrow 0$ and $N\rightarrow\infty$ with $T$
constant), the field remains in $\ket{0}$ \cite{Cpassisimple}.

An interesting insight into the Zeno effect on light is provided by considering
the back action of the measurement on the field phase. Each photon number
determination erases phase information, randomizing at each step the field
phase. The field amplitude undergoes a two-dimensional random walk, with a step
size $|\lambda| \Delta t$, instead of a deterministic addition of identical
displacements along a fixed direction. The final r.m.s. amplitude,
$\sqrt{N}|\lambda|\Delta t$, corresponding to $\overline{n}=|\lambda|^2 T
\Delta t$, coincides with the prediction
 based on the projection postulate.

In the analyzis of the Zeno effect, it is essential that the probability for
finding the system in its initial state has a quadratic evolution at short
times. There is no Zeno effect when the cavity is coupled to an environment
with a short memory time  $\tau_c$, inducing an evolution with probabilities
varying linearly with time. Assume for instance that the cavity, prepared in
vacuum, relaxes within the damping time $T_c$ towards a thermal equilibrium
with a blackbody mean photon number $n_b$ [9]. The probability for finding the
cavity in $|0\rangle$ after an initial time interval  $\Delta t \gg \tau_c$ is
$p_0 \approx 1-(n_b/T_c) \Delta t$, provided $n_b \Delta t/T_c\ll 1$. The
probability for being in $|0\rangle$ after $N$ steps of length $\Delta t$, each
followed by a measurement, is then $p_0^N \approx  [1-(n_b/T_c) \Delta t]^N
\approx 1-n_bT/T_c$ (with $n_bT/T_c \ll 1$). The field energy grows linearly at
short times, with the same rate as without measurements.

Similarly, the exponential damping of a field stored in the cavity leads to
state probabilities varying  linearly with time, which are not modified by
repeated measurements. The Zeno effect thus does not affect the irreversible
damping rate $T_c$ of the cavity. Note that other irreversible systems may
exhibit at short, but experimentally accessible times, a non-exponential
behaviour with a quadratic initial evolution. Quantum Zeno and anti-Zeno
effects have been theoretically described \cite{kurizki} and observed
\cite{zenotun} for such irreversible processes.

\begin{figure}
\begin{center}
\includegraphics[width=0.45\textwidth]{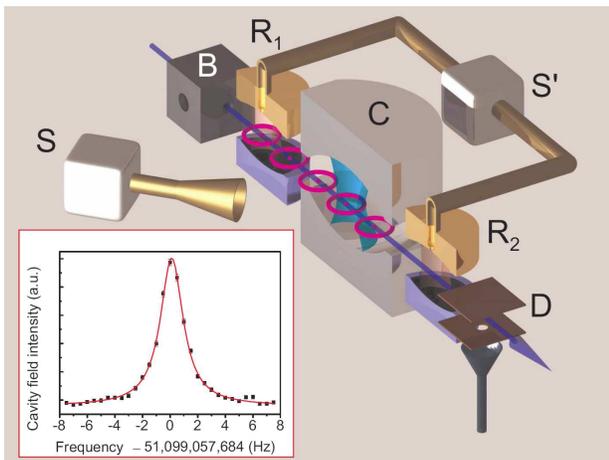}
\end{center}
\vspace{-0.3cm} \caption{\label{setup} Experimental set-up (cavities $C$, $R_1$
and $R_2$ are cut for clarity). The inset shows the cavity spectrum (dots) and
a lorentzian fit (line) with a 2.0~Hz FWHM.} \vspace{-0.5cm}
\end{figure}

Our experimental set-up \cite{exploring,rmp} is shown in figure \ref{setup}.
The key element is the superconducting cavity $C$, made up of two niobium
mirrors facing each other \cite{cavity}. It has a remarkably long damping time
$T_c=0.13$~s, when cooled to 0.8~K. The pulsed classical source $S$, tuned to
resonance with the cavity mode at 51.099~GHz, irradiates $C$ from the side. A
small fraction of the photons emitted in a pulse is coupled into $C$ through
diffraction on the mirrors edges. The other photons disappear
quasi-instantaneously when $S$ is switched off.

Phase stability is critical for the coherent accumulation of successive pulses
in $C$. An atomic clock stabilizes $S$ to a sub-Hertz frequency width and
stability. The lifetime-limited linewidth of $C$ is 1.2~Hz full width at half
maximum (FWHM). The cavity frequency must be stabilized to this level,
requiring a control of the 2.7~cm cavity length to a 0.5~picometer precision.
We stabilize the cavity length passively. The cryostat has been isolated from
external vibrations. We have stabilized the refrigerator temperature to $\pm
10^{-4}$~K and the pressure of the liquid helium bath to $\pm$0.1~mbar. Piezo
elements tune the mirrors separation and hence the mode frequency with a
2.4~kHz/V sensitivity. Their drive voltage (83~V) is stabilized to $\pm$0.1 mV
and drifts by less than $\pm$0.2 mV/h.

The inset in figure \ref{setup} shows the cavity resonance spectrum probed by
$S$ operating in a quasi-continuous emission mode. We switch off $S$ for short
time intervals, during which the field intensity in $C$ is measured by resonant
Rydberg atoms \cite{cavity}. These atoms are not submitted to the intense field
radiated outside $C$ when $S$ is on. The Lorentzian fit FWHM is 2.0~Hz,
differing from the expected 1.2~Hz width, due to a combination of short term
fluctuations and long term drift. During data acquisition, we measure the
cavity frequency every ten minutes and reset $S$ at resonance with $C$. If the
drift between two measurements is above 1.2~Hz, we eliminate the corresponding
data. We estimate that the $S-C$ detuning $\delta/2\pi$ is, on average,
0.6$\pm$0.2~Hz during data acquisition.

The non-resonant QND probe atoms are prepared in box $B$ from a
velocity-selected rubidium atomic beam (velocity $v=250\pm 1$~m/s) in the
circular Rydberg level $g$ (principal quantum number 50). Rydberg atom samples
are prepared within 2~$\mu$s pulses. Their position is well-known during their
flight through the apparatus. The cavity mode is close to resonance with the
transition from $g$ to $e$ (circular state with principal quantum number 51).
The atom-cavity detuning, $\Delta/2\pi=240$~kHz, is larger than the vacuum Rabi
frequency, $\Omega_0/2\pi=50$~kHz, which defines the atom-cavity coupling. The
atoms are transparent probes, unable to absorb or emit photons in $C$. The
$g\rightarrow e$ transition frequency is, however, light-shifted. This shift,
proportional to the photon number $n$, is used to measure it non-destructively
\cite{lifedeath,collapse}.

Before entering $C$, each atom is prepared in the superposition
$(\ket{e}+\ket{g})/\sqrt{2}$ by a $\pi/2$-pulse, resonant on the $g\rightarrow
e$ transition, produced by the classical source $S'$ in the low-$Q$ cavity
$R_1$. Using the Bloch pseudo-spin representation, we describe this
superposition as a spin in the $OX$ direction of the Bloch sphere. The light
shift accumulated during the atom-cavity interaction changes the phase of the
$e/g$ superposition, in a frame rotating at the atomic transition frequency in
vacuum \cite{collapse, exploring}. When $n$ photons are stored in $C$, the
atomic spin rotates by an angle $n\Phi_0$ in the equatorial plane of the
sphere. With the chosen values of $\Delta$ and $v$, the phase shift per photon
is $\Phi_0\approx\pi/4$. We measure the spin direction by applying on the atom,
at the exit of $C$, a $\pi/2$-pulse in $R_2$ with a phase $\varphi$ relative to
that of the $R_1$ pulse. The subsequent detection of the atomic state ($e$ or
$g$) in the field-ionization detector $D$ is equivalent to a spin detection at
the exit of $C$ along the $Ou$ axis in the equatorial plane, at an angle
$\varphi$ with $OX$.

\begin{figure}
\begin{center}
\includegraphics[width=0.45\textwidth]{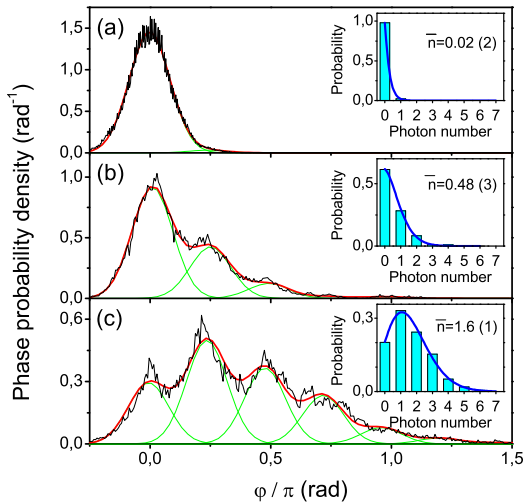}
\end{center}
\vspace{-0.3cm} \caption{\label{histos} Probability distributions of the atomic
spin phase after 0 (a), 20 (b) and 50 (c) field injections in $C$ (no
intermediate QND measurements). The thick red line is a fit to a sum of
gaussian distributions (thin green lines). Resolved peaks are centered at the
quantized light shifts, clearly illustrating field energy quantization. The
gaussian fits provide a direct measurement of the photon number probability
distributions shown in the insets, together with poissonian fits (solid lines).}
\end{figure}

The setting $\Phi_0\approx\pi/4$ is appropriate for a QND measurement up to
$n=7$ \cite{collapse}. The photon numbers $n=0\ldots 7$ are associated to 8
directions of the atomic spin at the exit of $C$. We send an ensemble of probe
atoms in a time shorter than $T_c$. For a given photon number $n$, all atomic
spins in this ensemble point in the same direction. We infer it by a quantum
tomography process, in which we detect the spins individually along one out of
four different axes corresponding to $\varphi/\pi=-0.250,\ -0.047,\ +0.247$ and
$+0.547$. With 110 atoms in the ensemble, the statistical noise on the
tomography process is of the order of the phase shift $\Phi_0$.

Let us first examine the coherent field growth without QND probe atoms between
the injections. An experimental sequence starts by reseting the field to
$\ket{0}$ using a large number of absorbing atoms, prepared in $g$ and set at
resonance with $C$ via the Stark effect. We then inject $N$ identical pulses
(50~$\mu$s duration) in $C$, separated by a time interval $T_i=5.04$~ms. At the
end of the sequence, we perform a QND measurement of the cavity field.

Figure \ref{histos}(a-c) presents the probability distributions of the measured
spin phase $\varphi$ for $N=0$, 20 and 50 injection pulses. For each sequence,
we detect $\approx$200 atoms over a $T_m=72$~ms duration, starting $2T_i$ after
the last injection. From these data, we extract about 90 non-independent
$\varphi$ values based on spin tomographies performed on a sliding window
containing an ensemble of 110 atomic detections. We repeat the procedure 500 to
2000 times. For $N=0$, the phase distribution is peaked at $\varphi=0$
revealing an empty cavity. For $N=20$, the peaks centered at $\Phi_0$ and
$2\Phi_0$ reveal the build-up of small probabilities for having one and two
photons. For $N=50$ injections, equidistant peaks are visible, corresponding to
$n=0\ldots 5$. These discrete peaks directly reveal the field energy
quantization in $C$ \cite{schoelkopf}. The phase distributions are fitted by
sums (thick red line) of Gaussians (thin green lines) with a unique standard
deviation of 0.266~$\pi$, corresponding to the spin tomography statistical
noise. From these fits, we extract $\Phi_0=$0.233~$\pi$. We also obtain the
photon number distributions and their averages. These distributions are
poissonian as expected for coherent states, see insets in figure \ref{histos}.

\begin{figure}
\begin{center}
\includegraphics[width=0.45\textwidth]{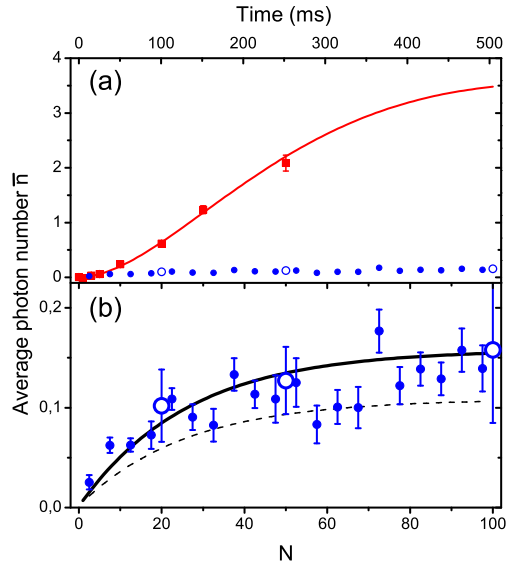}
\end{center}
\vspace{-0.3cm} \caption{\label{zenon} (a) Average photon number
$\overline{n}$ in $C$ as a function of the number $N$ of injection pulses
(bottom axis). The top axis gives a time scale for the complete sequence,
including injection and measurement times. The red squares correspond to the
uninhibited field growth, without measurements between injections (statistical
error bars for $N<50$ are smaller than point size). The initial evolution is
quadratic. The line is a fit providing a precise calibration of the injection
pulse amplitude. The blue dots result from atomic spin tomography performed between
injection pulses. The field growth is inhibited by the quantum Zeno effect. The
open circles correspond to QND measurements of the photon number using 200
atomic detections at the end of sequences with 20, 50 and 100
injection/measurement cycles. (b) Inhibited field evolution with an expanded
vertical scale (blue dots in a). The error bars reflect the statistical noise.
The solid and dotted black lines are theoretical predictions (see text). The
initial growth is now linear.}
\end{figure}

Cavity relaxation cannot be totally neglected during the delay between the last
injection and the measurement. We thus correct the measured mean photon number
by a scaling factor $1.4$, taking into account the $\approx 2T_i+T_m/2$ delay
between the last injection pulse and the center of the measurement time
interval. Figure \ref{zenon}(a) shows the evolution of the rescaled average
photon number $\overline{n}$ as a function of $N$ (red squares). Phase
stability problems prevented us from recording data for $N>50$. As expected,
the initial growth is quadratic. For large $N$ values, the sequence duration is
comparable to $T_c$ and $1/\delta$. For a comparison with theory, we must take
into account cavity damping and cavity-source detuning. The expected cavity
field amplitude after $k$ injections is:
\begin{equation}
\alpha_{k}=e^{-T_i/2T_c}\alpha_{k-1}+e^{i(k-1)T_i\delta}\alpha_1=\frac{e^{-kT_i/2T_c}-e^{ik\delta
T_i}}{e^{-T_i/2T_c}-e^{i\delta T_i}}\alpha_1\ ,
\end{equation}
where $\alpha_1$ is the first pulse amplitude, which is taken as the phase
reference ($\alpha_1$ assumed to be real, without loss of generality). The
value of $\alpha_1$ is deduced from a fit of the calculated photon number with
the data (solid line in figure \ref{zenon}(a)), taking into account the
estimated $\delta$ value. We get $\overline{n}_1=\alpha_1^2=0.00223\pm0.00012$.

We now send QND probe atoms between injections and record their final state
thus acquiring information from each realization of the field state. In order
to determine the evolution of the average photon number, we perform a
statistical analyzis of many such realizations, using a simplified version of
the spin tomography method described above. Since the photon number is now
expected to remain small, we use a single detection phase,
$\varphi=-0.278~\pi$, which optimally discriminates the spin directions
corresponding to 0, 1 and 2 photons. To obtain the spin histogram and thus
determine the average photon number at a given time, we collect the information
from atoms detected in a 25 ms window around this time ($\sim 50$ atoms
detected during 5 injection/measurement cycles) and repeat the procedure from
500 to 2500 times. The basic conclusions of our initial qualitative discussion
are valid, since the amplitude injected by five consecutive pulses is still
very small.

The corresponding average photon number, $\overline{n}$, is shown (blue dots)
as a function of $N$ in figure \ref{zenon}(a). All $\overline{n}$ values are
determined from independent samples made of 50 detected atoms. We check this
simplified measurement by a complete spin tomography performed on $\approx$200
detected atoms at the end of sequences with 20, 50 and 100
injection/measurement cycles (open circles in figure \ref{zenon}(a)). The field
growth is almost completely inhibited, $\overline{n}$ remaining below 0.2.

Figure \ref{zenon}(b) presents the inhibited field growth with an expanded
vertical scale. This growth can be modelled by a random walk, assuming that the
field phase is blurred between two injections. The dotted black line in figure
\ref{zenon}(b) presents the prediction of this model with no adjustable
parameter. This prediction is slightly below the observations, indicating that
phase is not totally blurred between injections. We have performed a detailed
quantum Monte Carlo simulation of the field phase diffusion induced by the
atoms crossing $C$. It includes the cavity relaxation, the finite 0.8~K
temperature of its mode ($n_b=0.05$ blackbody photons on average) and the
finite detection efficiency of $D$ (50\%). The simulation results (solid black
line in figure \ref{zenon}(b)) are in excellent agreement with the observations
and clearly exhibit the initial linear growth of $\overline{n}$, characteristic
of the Zeno effect.

We have observed a clear manifestation of the quantum Zeno effect on light.
Repeated intensity measurements performed on the field in a high-$Q$ cavity
inhibit the quadratic run-away of the energy fed by a pulsed classical source.
The residual field growth is modelled by a random walk in phase space. This
effect illustrates vividly the back action of the photon number measurement
onto the field phase. The realization of this experiment opens interesting
perspectives for controlling quantum systems. Instead of freezing their
evolution, repeated measurements could provide information used to channel them
towards tailored quantum states by active feed-back operations
\cite{feedbackfock}.

{\bf Acknowledgements} This work was supported by the Agence Nationale pour la
Recherche (ANR), by the Japan Science and Technology Agency (JST), and by the
EU under the IP project SCALA. S.D. is funded by the D\'el\'egation
G\'en\'erale \`a l'Armement (DGA).


\begin{thebibliography}{99}

\vspace{-0.5cm}

\bibitem{zenotheo}
B. Misra and E. C. G. Sudarshan, J. Math. Phys. Sci. \textbf{18}, 756 (1977);
K. Koshino and A. Shimizu, Phys. Rep. \textbf{412}, 191 (2005) and references
therein.

\bibitem{Home}
D. Home and A. A. B. Whitaker, Annals of Physics \textbf{258}, 237 (1997).

\bibitem{Rauch}
H. Nakazato M. Namiki, S. Pascazio and H. Rauch, Phys. lett. A \textbf{217},
203 (1996).

\bibitem{zenomolec}
B. Nagels, L. J. F. Hermans and P. L. Chapovsky, Phys. Rev. Lett. \textbf{79},
3097 (1997).

\bibitem{zenoion}
W. M. Itano, D. J. Heinzen, J. J. Bollinger and D. J. Wineland, Phys. Rev. A
\textbf{41}, 2295 (1990); Chr. Balzer, R. Huesman, W. Neuhauser and P. Toschek,
Opt. Comm. \textbf{180}, 115 (2000).

\bibitem{zenobec}
E. W. Streed \textit{et al.}, Phys. Rev. Lett. \textbf{97}, 260402 (2006).

\bibitem{zenopolar}
P. G. Kwiat \textit{et al.}, Phys. Rev. Lett. \textbf{83}, 4725 (1999); O.
Hosten \textit{et al.}, Nature \textbf{439}, 949 (2006).

\bibitem{lifedeath}
S. Gleyzes \textit{et al.}, Nature \textbf{446}, 297 (2007).

\bibitem{collapse}
C. Guerlin \textit{et al.}, Nature \textbf{448}, 889 (2007).

\bibitem{exploring}
S. Haroche and J. M. Raimond, \textit{Exploring the Quantum} (Oxford University
Press, Oxford, 2006).

\bibitem{Cpassisimple}
This analysis of the Zeno effect overlooks events in which the field ends up in
vacuum after beeing transiently excited. For $N \rightarrow \infty$ the
probability of such events, which involves terms of order $|\lambda|^4 \Delta
t^4$, is negligibly small. See [3] for an analyzis of a similar problem in the
Zeno effect of a two-level system.

\bibitem{kurizki}
A. G. Kofman and G. Kurizki, Nature \textbf{405}, 546 (2000); P. Facchi, H.
Nakazato, and S. Pascazio, Phys. Rev. Lett. \textbf{86}, 2699 (2001).

\bibitem{zenotun}
M. C. Fischer, B. Guti\'errez-Medina and M. G. Raizen, Phys. Rev. Lett.
\textbf{87}, 040402 (2001).

\bibitem{rmp}
J. M. Raimond, M. Brune, and S. Haroche, Rev. Mod. Phys. \textbf{73}, 565
(2001).

\bibitem{cavity}
S. Kuhr \textit{et al.}, Appl. Phys. Lett. \textbf{90}, 164101 (2007).

\bibitem{schoelkopf}
D. I. Schuster \textit{et al.}, Nature \textbf{445}, 515 (2007).

\bibitem{feedbackfock}
J. M. Geremia, Phys. Rev. Lett. \textbf{97}, 073601 (2006).



\end{thebibliography}
\end{document}